\documentstyle[twoside,fleqn,espcrc2,epsf]{article}

\title{First results from a parametrized Fixed-Point QCD action\thanks{Talk
    presented by S.\ Hauswirth}}
\author{P.\ Hasenfratz\address[BERN]{Institute for Theoretical Physics,
    University of Bern, Sidlerstrasse 5, CH-3012 Bern,
    Switzerland}, 
    S.\ Hauswirth\addressmark, K.\ Holland\addressmark, 
    T.\ J\"org\addressmark, F.\ Niedermayer\addressmark}

\begin{document}

\begin{abstract}
We have constructed a new fermion action which is an approximation to the (chirally symmetric) Fixed-Point action, containing the full Clifford algebra
      with couplings inside a hypercube and paths built from renormalization
      group inspired fat links. We present an exploratory study of the light
      hadron 
      spectrum and the energy-momentum dispersion relation. \vspace{0mm}
\end{abstract}

\maketitle

\section{Introduction}
Fixed-Point (FP) fermions are a
solution to the Ginsparg-Wilson equation \cite{Hasenfratz:1998ft} and
thus preserve chiral 
symmetry on the lattice \cite{Luscher:1998pq}. While there exist a range of lattice
fermion formulations with exact or approximate chiral symmetry \cite{Narayanan:1993ss,Kap92,Gattringer:2001qu,DeGrand:2001tf,Bietenholz:2000iy}, we
expect FP fermions to have additional improved properties like 
small scaling violations. For
the FP gauge 
action we work with, good scaling behaviour has been found for
measurements of $T_c$ and glueballs up to lattice spacings of $\sim$0.3
fm \cite{Niedermayer:2001yx}. 
We report first
results for light hadron spectroscopy with a hypercubic FP fermion action
\cite{Hasenfratz:2001qb} which includes all Clifford algebra elements and
uses gauge paths constructed out of fat links. Results for
the chiral condensate and other observables are given in
\cite{Hasenfratz:2002}.

\section{Simulation parameters}

At a gauge coupling of $\beta=3.0$, the lattice spacing
determined from 
$r_0$ is $a=0.16$ fm. Lattice
volumes of $6^3 \times 16$ and $9^3\times 24$ are compared to identify finite
volume effects. We use gaussian smeared sources and
point sinks. Configurations are fixed to Landau gauge.
Bare quark masses  $m_qa$ range from 0.016 to 
0.23. At the smallest mass we get $m_\pi/m_\rho = 0.33(3)$. 
In this exploratory study we introduced the mass by defining $D(m_q) = D^{\rm
  FP} + m_q$. 
A multimass BiCGStab algorithm 
is used to invert the Dirac operator.  
We symmetrize the correlators, make correlated fits and
calculate bootstrap errors. More details on the simulation and results can be
found in \cite{Hasenfratz:00wip}.

\begin{figure}[htb]
\epsfxsize=\hsize
\epsfbox{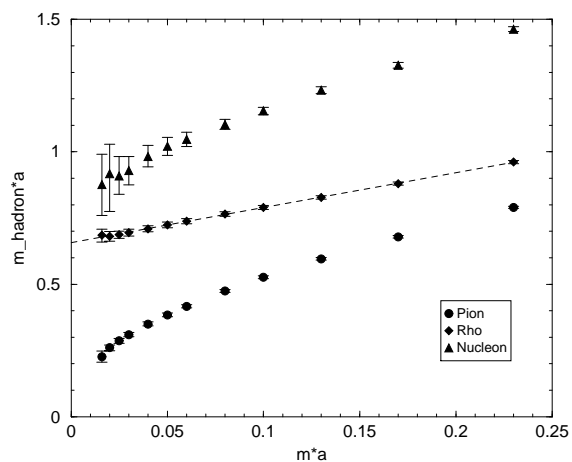}
\vspace{-9mm}
\caption{$\pi$, $\rho$ and $N$ masses for $D^{\rm FP}$ at $\beta=3.0$ from 70
  $9^3\times 24$ gauge configurations.}
\vspace{-4mm}
\label{fig:fp_spectrum}
\end{figure}

\begin{figure}[htb]
\epsfxsize=\hsize
\epsfbox{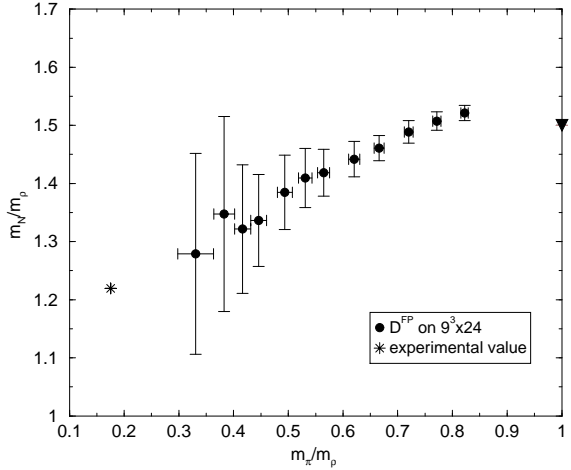}
\vspace{-9mm}
\caption{Edinburgh plot for $D^{\rm FP}$.}
\vspace{-4mm}
\label{fig:fp_edinburgh}
\end{figure}

Fig.~\ref{fig:fp_spectrum} shows the masses of the $\pi$, $\rho$ and $N$
measured with the parametrized FP Dirac operator $D^{\rm FP}$ on the
$9^3\times 24$ lattice. The inversion 
of the Dirac operator converged for all configurations,
although for one configuration the number of iterations was about $2.5$ times
larger than the typical value. This indicates that the
residual chiral symmetry breaking leads to fluctuations of the order of the
smallest mass for the real eigenvalues of $D^{\rm FP}$. The scale determined
from the rho 
mass is 
$a=0.17$ fm. Fig.~\ref{fig:fp_edinburgh} 
shows an Edinburgh plot of the data.

\section{Spectroscopy with $D^{\rm FP}$}
We use three different correlators to extract the pion mass: The pseudoscalar
$P$, the axial vector $A$
and the difference of pseudoscalar and
scalar $P$-$S$. In the quenched theory, the $P$
correlator picks up a contribution proportional to
$Q/m_q^2V$. This quenched enhanced finite size effect is
cancelled in the difference $P$-$S$ 
if the action is exactly
chiral. For the $A$ correlator, the divergence is
expected to be partially 
suppressed \cite{Blum:2000kn}.

\begin{figure}[htb]
\epsfxsize=\hsize
\epsfbox{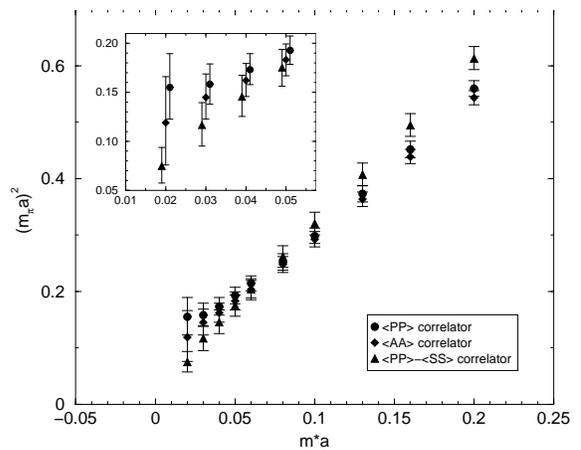}
\vspace{-9mm}
\caption{$m_{\pi}^2$ for the parametrized FP Dirac
  operator $D^{\rm FP}$ from 100 $6^3\times 16$ gauge
  configurations.}
\label{fig:fp_pisquared_6}
\end{figure}

\begin{figure}[htb]
\epsfxsize=\hsize
\epsfbox{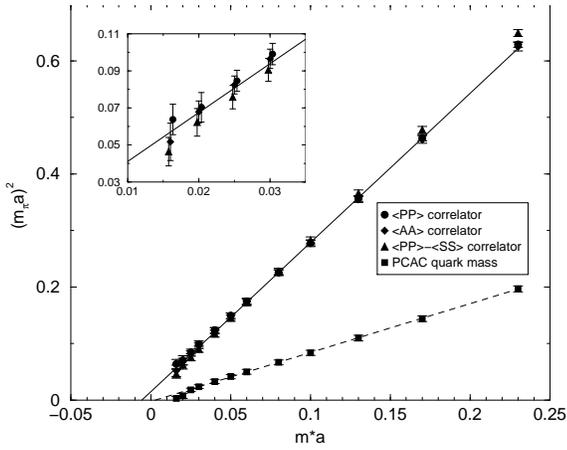}
\vspace{-9mm}
\caption{$m_{\pi}^2$ and the PCAC
  quark mass for $D^{\rm FP}$ from 70 $9^3\times 24$ gauge
  configurations.}
\vspace{-4mm}
\label{fig:fp_pisquared_9}
\end{figure}

Figs.~\ref{fig:fp_pisquared_6} and \ref{fig:fp_pisquared_9} show a comparison
of $m_{\pi}^2$ 
vs.~$m_q$ at the two lattice sizes using the parametrized FP Dirac operator. On the smaller volume, the three different pion
correlators give different values at small quark masses. The $P$ correlator
lies highest, while the $P$-$S$ correlator gives considerably smaller pion
masses. The $A$ correlator
lies in between. This behaviour is in qualitative agreement with the
results from domain-wall fermions \cite{Blum:2000kn}. For larger $m_q$,
the $P$-$S$ correlator deviates from the pion due to the
the closely lying scalar state. 
 The discrepancy between the three 
correlators in the chiral limit is reduced at our
larger lattice volume, where the topological
finite-volume effects are on the order of 
the statistical error. 

On the $9^3 \times 24$ lattice, we also measure the unrenormalized PCAC quark
mass 
$$
2m^{\rm PCAC}(t) = \sum_{\vec x} \langle\partial_4 A_4(\vec x, t)P(0)\rangle / \sum_{\vec
 x} \langle P(\vec x, t)P(0)\rangle, 
$$
which should go to zero as $m_q\rightarrow 0$ if the Dirac operator respects
chiral symmetry. From a linear fit to all but
the smallest two masses we get $m_{\rm res}^{\rm PCAC}a=0.003(4)$.
A linear fit to $m_{\pi}^2$ gives $m_{\rm res}a=-0.007(3)$,
$-0.006(2)$ and $-0.002(3)$ for the
$P$,  $A$ and $P$-$S$ correlators, respectively. Hence we do not see
additive mass renormalization due to residual chiral symmetry breaking within
our 
statistical and systematical errors.

\begin{figure}[htb]
\epsfxsize=\hsize
\epsfbox{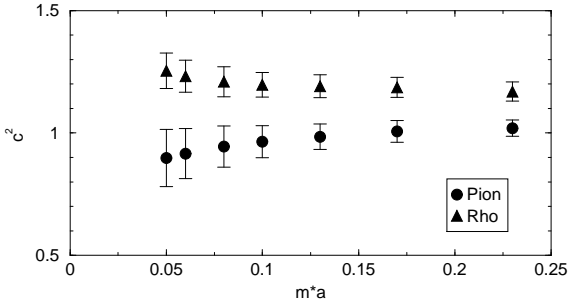}
\vspace{-9mm}
\caption{$c^2$ for pion and rho meson
  with $D^{\rm FP}$.}
\vspace{-4mm}
\label{fig:c_squared}
\end{figure}

In Fig.~\ref{fig:c_squared} the squared speed of light $c^2 = 
(E(p)^2 - m^2)/p^2$ measured on the $9^3\times 24$ lattice is shown for the
smallest momentum $|\vec p| = 2 \pi /9$. While $c^2$ for the pion is consistent with 1 within errors, it
is not for the rho meson. However, the error bars do not
include systematic uncertainities from choosing the fit range. The dispersion
relation 
is significantly improved compared to the Wilson or clover action.

\section{Overlap improved FP Dirac operator}

\begin{figure}[htb]
\epsfxsize=\hsize
\vspace{-1.5mm}
\epsfbox{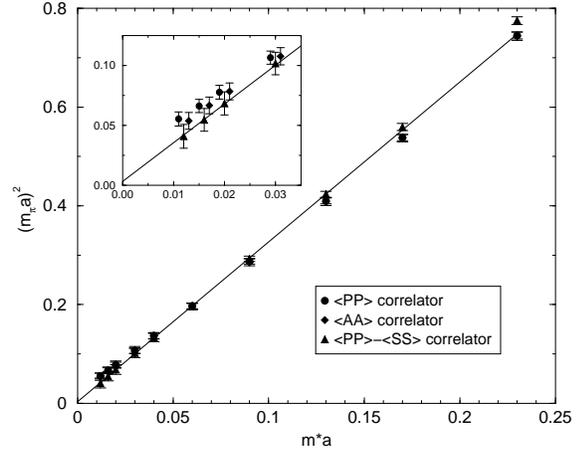}
\vspace{-9mm}
\caption{$m_{\pi}^2$ for $D^{\rm FP}_{\rm ov}$ from 28 $9^3\times 24$ gauge
  configurations at $\beta=3.0$.}
\vspace{-4mm}
\label{fig:sqrpion_overlap}
\end{figure}

We define the massive overlap Dirac operator
\begin{equation}
D^{\rm FP}_{\rm ov}(m_q) = (1-m_q/2) D^{\rm FP}_{\rm ov}(0) + m_q/2R,
\end{equation}
with $R$ from the Ginsparg-Wilson relation and 
\begin{equation}
D^{\rm FP}_{\rm ov}(0) = (2R)^{-1/2}  (1- A/\sqrt{A^\dag A}) (2R)^{-1/2}.
\end{equation}
In the kernel $A = 1 - \sqrt{2R} D^{\rm FP} \sqrt{2R}$ the pa\-ra\-me\-trized FP
Dirac operator enters.
We use a 3$^{\rm rd}\!$ order Legendre expansion of the inverse
square root and project out the smallest 20 eigenvalues of $A^\dag
A$.  At the smallest mass $m_qa=0.012$ we get
$m_\pi/m_\rho=0.27(4)$.
Fig.~\ref{fig:sqrpion_overlap} shows $m_\pi^2$ for the overlap
improved FP Dirac operator. A linear fit to the $P$-$S$-correlator at the six
smallest masses is consistent with zero at $m_q=0$.
Similar behaviour has been observed at $\beta=3.2$ ($a=0.13$ fm).

\section{Acknowledgements}

This work has been supported in part by the Schweizerischer
Nationalfonds and the European Community's Human Potential Programme under
HPRN-CT-2000-00145, Hadrons/Lattice QCD. We thank the Swiss Center
for Scientific Computing in Manno for computing resources and the Regensburg
group for useful discussions.

\end{document}